\documentclass[runningheads]{llncs}
\usepackage{graphicx}
\usepackage{amsfonts}
\usepackage{amsmath}
\usepackage{why3style}
\usepackage{multirow}
\usepackage{hyperref}
\hypersetup{
    colorlinks=true,
    linkcolor= black,
    filecolor= black,
    urlcolor=blue,
    citecolor = black,
    pdfpagemode=FullScreen,
    }
\usepackage{float}
\usepackage{xcolor}
\usepackage{colortbl}
\usepackage{rotating}

\begin{document}
\title{Leroy and Blazy were right: their memory model soundness proof is automatable\\(Extended Version)}
%
% \title{Mostly-automated formal verification of a \\C-like memory model in Why3}
\titlerunning{Leroy and Blazy were right}
\author{Pedro Barroso %\orcidID{0000-1111-2222-3333} 
\and
Mário Pereira%\orcidID{1111-2222-3333-4444} 
\and
António Ravara%\orcidID{2222--3333-4444-5555}
}
\authorrunning{P. Barroso, M. Pereira, A. Ravara}
\institute{NOVA School of Science and Technology \and NOVA-LINCS, Portugal}
\maketitle              % typeset the header of the contribution

\begin{abstract}

Xavier Leroy and Sandrine Blazy in 2007 conducted a formal verification, using the Coq proof assistant, of a memory model for low-level imperative languages such as C. Considering their formalization was performed essentially in first-order logic, one question left open by the authors was whether their proofs could be automated using a verification framework for first-order logic. 
We took the challenge and automated their formalization using Why3, significantly reducing the proof effort. We systematically followed the Coq proofs and realized that in many cases at around one third of the way Why3 was able to discharge all VCs.
Furthermore, the proofs still requiring interactions (e.g.\ induction, witnesses for existential proofs, assertions) were factorized isolating auxiliary results that we stated explicitly.
In this way, we achieved an almost-automatic soundness and safety proof of the memory model. Nonetheless, our development allows an extraction of a correct-by-construction concrete memory model, going thus further than the preliminary Why version of Leroy and Blazy.

%AR: REFER THAT WE DEFINED AN EXTRACTABLE CONCRETE MODEL, GOING THUS FURTHER THAN THE ORIGINAL WORK

%Lemas "explicitos" ajudam a compreender a prova, revela a essencia dos resultados/prova ao olhar apenas para o código

\end{abstract}
%\tableofcontents
%\newtheorem*{definition}{Definition}

\section{Introduction}

%\subsection{Context}~

Formal semantics are concerned with the process of building a mathematical model to serve as a basis for understanding and reasoning about how programs behave. A mathematical model is important because the activity of trying to precisely define the behavior of program constructions can reveal all types of subtleties of which it is crucial to be aware \cite{WinskelFormalSemantics}. Many programs that require formal verification are written in imperative languages that accommodate pointer-based data structures with in-place modifications. To reason about the contents of the memory or even the behavior of operations over it, one needs to develop an adequate memory model.

\begin{figure}[ht]
    \centering
    \includegraphics[width=0.65\textwidth]{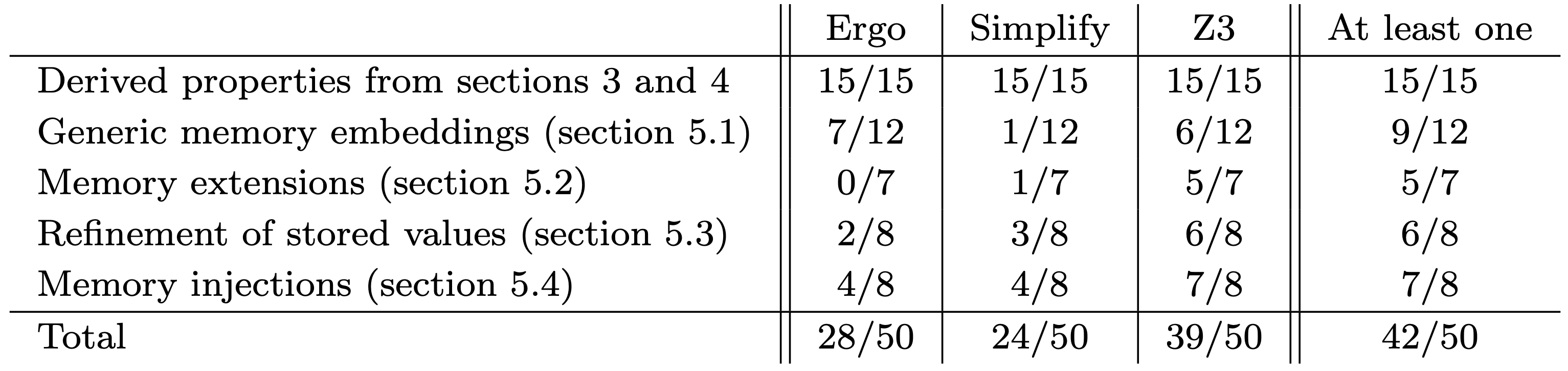}
    \caption{Leroy and Blazy's Why formalization}
    \label{fig:whytable}
\end{figure}

\vspace{-0.5em}
\paragraph{Leroy and Blazy}in 2007 formalized and verified, using the Coq proof assistant, a memory model for C-like imperative languages \cite{formalclikememorymodel}. Coq proofs tend to be a very time consuming task and in fact, Leroy and Blazy's proof was almost the length of the specification and theorems (970 and 1070 lines, respectively). Considering their formalization was performed essentially in first-order logic, one question the authors addressed was whether their proofs could be automated using a verification framework for first-order logic, e.g.\ Why3 \cite{filliatre2013why3}. The authors translated their memory model to Why (a former version of Why3), the authors claim that at the time some recursive definitions were hard to define in the Why syntax and therefore, they only translated the axiomatizations and derived properties of the memory model. Their preliminary study (Table \ref{fig:whytable}) showed that % most likely it can
could be automated, but no further study or complete formalization of their memory model has been automated so far.

\vspace{-1em}
\vspace{-1em}

\noindent Nonetheless, the problem we address here is: 

\begin{quote}\emph{
    Is it possible (15 years later) to automate the complete abstract and concrete implementation of the authors memory model? 
    }
\end{quote}

%Provas legiveis (transformações)
%Menos esforço de prova
%Nos casos que foram preciso interações foram significativamente menores que em Coq
%Automatização destas interações em Why3
%

\vspace{-2em}
\paragraph{In this paper}we translate Leroy and Blazy's memory model to Why3 and develop an almost-automatic soundness and safety proof. 
Moreover, our code, implementing a concrete memory model, is extractable. This requires some manual work, namely re-organizing modules, that we believe with some refactoring of the WhyML code could be avoided. \\

\vspace{-0.5em}
Concretely, our contributions are:
\begin{itemize}
    \item two Why3 versions (one containing interactive proofs and one ``fully'' automatic) that reduce up to 90\% the Coq proof effort;
    \item general techniques to achieve a more automatic proof; %explicitly state interactive transformations within the Why3 code;
    \item a proof effort comparison between %both Why3
    ours and the original Coq ones;
    \item a correct-by-construction memory model extracted from our development.
\end{itemize}
%
%Why3 \cite{filliatre2013why3} is a platform for deductive program verification based on typed first-order logic that concedes a high degree of automation, which is beneficial in a pedagogical context. It uses WhyML, an ML-style programming language with extensive support for specification annotations. To prove a program, Why3 uses a standard weakest-precondition procedure \cite{weakestpreconditions} to generate verification conditions (VC) and then discharge them using third-party theorem provers. Why3 supports a vast of third-party theorem provers (automated and interactive), so VC can be discharged using a combination of provers rather than a single one.
%
% \begin{itemize}
%     \item Memory model and why it is important
%     \item Compcert and Leroy and Blazy memory model
% \end{itemize}
%
% \subsection{Problem}
%
% \begin{itemize}
%     \item Coq proof effort
%     \item Too exhaustive
%     \item Time consuming
%     \item "Little" effort to translate it to Why3 since both share several similarities
%     \item Coq more rigorous (have almost 100\% certainty) vs relying on SMT solvers, where to draw the line?
% \end{itemize}
%
% \subsection{Contributions}
%
%Provas legiveis (transformações)
%Menos esforço de prova
%Nos casos que foram preciso interações foram significativamente menores que em Coq
%Automatização destas interações em Why3
%
%
\vspace{-2em}
\paragraph{The rest of the article}is organized as follows. Section \ref{sec:memorymodel} describes %the importance of a 
memory models and briefly presents the one of Leroy and Blazy. Also, Section \ref{sec:memorymodel} shows how to prove the memory model sound and safe (which includes semantics preservation for three passes of the Compcert compiler), and the Coq proof statistics that Leroy and Blazy stated in their paper. 
Section \ref{sec:why3} outlines our Why3 development and the techniques used to explicitly state interactive transformations. Section \ref{sec:proofeffort} compares both versions with the Coq version and presents our approach to extract correct-by-construction OCaml code, followed by conclusions in Section~\ref{sec:conclusions}.  % which in terms of specification is roughly the same as the Coq one. However, Why3 is able to prove automatically 92.7\% of it. 
Our Why3 development and OCaml extracted code is publicly available\footnote{\url{https://gitlab.com/p.barroso/memory-model-c-why3/}}.
\vspace{-1em}
\section{The memory model}
\label{sec:memorymodel}
Memory models provide the necessary abstraction between the behavior of a program and the behavior of the memory it reads and writes. 
When reasoning about compilers and low-level code we need to account for several factors such as memory management, concurrency behavior, casts, structured pointers, overlapping locations, etc. Therefore, it is not sufficient to interpret memory as a assignment of values to locations. 

\vspace{-1em}
\subsection{Concept} %Overview (pensar)

\begin{figure}[ht]
\centering
\includegraphics[width=8cm]{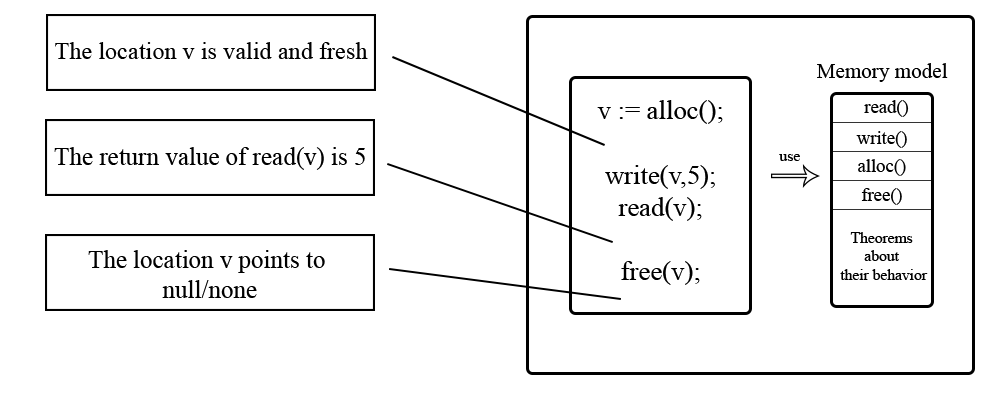}
\caption{Concrete assertion over a simple program}
\label{fig:concreteassert}
\end{figure}

The generic idea behind a memory model is to explicitly describe and provide some guarantees on the behavior of certain operations that manipulate memory (e.g.\ read, write, allocation, free). This allows to make concrete assumptions about the ``state of the memory'' at any time of the execution of a program. One example is that reading after writing a value into a location should return the value that was previously stored. Figure \ref{fig:concreteassert} shows this simple scenario and the natural assertions one can make over the execution of the program.

%intercalar
There are several ways to define a memory model: as a description of the set of valid traces of operations fulfilled by the execution of a program (e.g.\ IBM POWER multiprocessors \cite{ibmmulti}); as an abstract machine that receives and replies to messages (e.g.\ CompCertTSO \cite{compcerttso}); and as a set of functions that can be invoked along with some guarantees on their results (e.g.\ Leroy and Blazy's memory model \cite{formalclikememorymodel}).
Note that in most cases a definition in one of these styles is provably equivalent to a definition in another style \cite{Mansky_2015}.

%The first approach is mostly taken in concurrent memory models (e.g.\ IBM POWER multiprocessors \cite{ibmmulti}), where one describes a set of rules that allow or forbid some sequences of memory operations. In the second, memory is a whole different component and separate from the program with its own transition system, where system progress is produced by combining program steps and memory steps (e.g.\ CompCertTSO \cite{compcerttso}). In the last, which is the style Leroy and Blazy followed, each operation is a function with its own arguments and return type, and the results of the functions have specific restrictions.

The memory model presented by Leroy and Blazy is used in the formal verification of the Compcert \cite{Leroy_2009} compiler, which transforms the \texttt{Clight} subset of the C programming language down to \texttt{PowerPC} assembly code.

\vspace{-0.5em}
\subsection{Leroy and Blazy's memory model}
\label{mem:axioms}

Leroy and Blazy start by giving an abstract, incomplete specification of a memory model %that tries 
to formalize the memory-related aspects of C and related languages. The abstract type \texttt{val} represents values, which  includes the constant \texttt{vundef} to describe an undefined value. 
To describe references to memory blocks they use an abstract type \texttt{block} and to represent memory states an abstract type \texttt{mem}. 

\vspace{-0.5em}

\paragraph{Memory definitions.}
A memory \emph{state} is a collection of separated blocks where each block behaves as an array of bytes, and is addressed using byte offsets $i \in \mathbb{Z}$.
A memory \emph{location} is a pair (\texttt{b}, \texttt{i}) of a block reference \texttt{b} and an offset \texttt{i} within this block.
Lastly, the constant \texttt{empty}~:~\texttt{mem} represents the initial memory state.

\vspace{-1em}
\subsubsection{Operations.}
They define four operations that manipulate memory states as total functions:
\vspace{-1em}
\begin{alignat*}{2}
    \texttt{alloc} &: \texttt{mem} \times \mathbb{Z} \times \mathbb{Z} \rightarrow \texttt{option(block} \times \texttt{mem)} \\
    \texttt{free} &:  \texttt{mem} \times  \texttt{block} \rightarrow  \texttt{option mem} \\
    \texttt{load} &:  \texttt{memtype} \times  \texttt{mem} \times  \texttt{block} \times Z \rightarrow  \texttt{option val} \\ 
    \texttt{store} &: \texttt{memtype} \times  \texttt{mem} \times  \texttt{block} \times Z \times  \texttt{val} \rightarrow  \texttt{option mem}
\end{alignat*}
All these functions return option types to take into account potential failures. The function \texttt{alloc(m,l,h)} allocates a fresh memory block, where \texttt{m} is the initial memory state, $\texttt{l} \in \mathbb{Z}$ is the lower bound of the block (inclusive), and $\texttt{h} \in \mathbb{Z}$ the upper bound (exclusive) of the fresh block. Allocation can fail if not enough memory is available. Otherwise, \texttt{Some (b,m)} is returned, where \texttt{b} is the reference to the new block and \texttt{m} the updated memory state.

Contrarily, \texttt{free(m,b)} deallocates block \texttt{b} in memory \texttt{m}. In case of success, an updated memory state is returned.

The function \texttt{store(}$\tau$\texttt{,m,b,i,v)} writes value \texttt{v} of type $\tau$ at offset \texttt{i} in block~\texttt{b} of~\texttt{m}. If successful, the updated memory state is returned.

Symmetrically, \texttt{load(}$\tau$\texttt{,m,b,i)} reads a data type of $\tau$ from block \texttt{b} of memory state \texttt{m} at byte offset \texttt{i}. If successful, the value from block \texttt{b} is returned. %The memory state remains the same.

\vspace{-1em}
\subsubsection{Axiomatization and additional properties.} 
The authors axiomatize the expected behavior of the operations.\footnote{The complete set of hypotheses can be consulted \href{https://xavierleroy.org/memory-model/}{online}.} Recall that any implementation of the model must satisfy all these properties.

\vspace{-0.5em}
\paragraph{Good and not so good variables (axioms S5 to S8).} 
Define the correct behavior (operation succeeds) of a \texttt{load} after an \texttt{alloc}, \texttt{free} or \texttt{store} operation. Concretely, these hypotheses specify that: \texttt{alloc} and \texttt{free} preserve \texttt{load}s performed in any other disjoint block; reading from the same location with a compatible type \texttt{$\tau'$} succeeds and returns the value \texttt{convert(v,$\tau'$)}; and storing a value of type \texttt{$\tau'$} in block \texttt{b} at offset \texttt{i} commutes with loading. Furthermore, Leroy and Blazy also specify when the \texttt{load} operation returns an undefined~value.  

\vspace{-0.5em}
\paragraph{Block validity (axioms S9 to S13).}
The correct behavior of a \texttt{load} after an \texttt{alloc}, \texttt{free} or \texttt{store} depends on separation properties between blocks. In order to capture such properties, Leroy and Blazy axiomatize the relation \texttt{m} \texttt{$\models$} \texttt{b}, which means that block~\texttt{b} is valid in memory \texttt{m}. A block is valid if it was previously allocated and was not yet deallocated.

\vspace{-0.5em}
\paragraph{Bounds of block (axioms S14 to S17).}
The authors also axiomatize the function \texttt{$\mathcal{B}$(m, b)} that associates low and high bounds to a block \texttt{b} in memory state \texttt{m}. The axiomatization specifies that a freshly allocated block has the bounds that were given as argument to the \texttt{alloc} function and the bounds of a block are preserved by an \texttt{alloc}, \texttt{store} or \texttt{free} operation over a different block.

\vspace{-0.5em}
\paragraph{Valid access (axiom S18 and derived properties D19 to D22).}
Combining the definitions of block validity and of bounds of a block, Leroy and Blazy define the ``valid access'' relation \texttt{m $\models$ $\tau$ @ b, i}, which means that in state~\texttt{m}, the block \texttt{b} is a valid block and the range of byte offsets being accessed is included in the bounds of \texttt{b}. Intuitively, the \texttt{store} and \texttt{load} operations succeeds if and only if the corresponding memory reference is valid.

\vspace{-0.5em}
\paragraph{Freshness property (properties P30 to P34).} In Leroy and Blazy's concrete memory model, \texttt{alloc} never reuses block identifiers. Therefore, they define the relation \texttt{m \# b} stating that block \texttt{b} is fresh in memory \texttt{m}. This relation is mutually exclusive with block validity.

\vspace{-0.5em}
\paragraph{Alloc determinism (property P35).} Lastly, Leroy and Blazy state that alloc is deterministic with respect to the domain of the current memory state, i.e.\, alloc chooses the same free block when applied twice to memory states that have the same domain, but may differ in block contents.
\vspace{-0.5em}
\subsection{Compiler passes and their soundness proof}

%\begin{itemize}
%    \item Precise Handling of Undefined Values
%    \item Memory Allocation Behavior
%    \item Good Variable Properties
%    \item Memory transformations
%    \begin{itemize}
%    \item Generic memory embeddings
%    \item Memory extensions
%    \item Refinement of stored values
%    \item Memory injections
%    \end{itemize}
%\end{itemize}

The axiomatization of the previous section specify precise handling of undefined values, behavior of operations over memory states and ``good-variable'' properties. %We now use t
This theory is used to prove the correctness of program transformations performed by three passes of the \texttt{Compcert} compiler. Our approach follows the same process presented by Leroy and Blazy and, therefore, we also prove the correctness of these transformations. %Let us briefly introduce the three passes.

%The first pass is the translation from the source language \texttt{Clight} to the intermediate language \texttt{Cminor}. %On one hand, the Clight program allocates N fresh blocks at each function entry for its N local variables. On the other, the Cminor allocates only one.
%The second pass is register allocation, where local variables and pseudo-registers (temporaries) are initialized to the \texttt{undef} value on function entry and after register allocation, some of these variables and temporaries are mapped to global hardware registers. %The original program could have stored values of uninitialized variables in memory locations, henceforth, the memory state before and after register allocation have the same shapes but the contents may differ.
%The last is the spilling pass performed after register allocation, %The spilling
%which enlarges the stack frame that was defined at the time of \texttt{Cminor} generation, to make room for variables variables and temporaries that could not be allocated to hardware registers.

%The memory behaves differently between the source Clight program and the transformed Cminor.  (load and store operations performed by Clight either disappear in Cminor or becomes load and stores in sub-areas of the Cminor stack block).

To prove soundness and safety of each transformation, one need to guarantee specific results between the memory operations performed by the original and transformed program. For that, invariants are defined to describe the memory states at every point of the execution of the original and transformed programs. 

Leroy and Blazy use four relations between memory states: memory embeddings; memory extensions; refinement of stored values; and memory injections. For each of them, they define several properties the program needs to hold in order to prove soundness and safety. We decided to omit the exact definitions, %in this paper, 
as the list is exhaustive and falls out of the scope of this paper.\footnote{The authors did not presented a summarized list of properties for each of the four relations.} 

\vspace{-0.5em}
\subsection{Original Coq proof}

Coq uses \textit{Gallina}, a functional programming language. The close connection between functional languages and mathematical definitions eases the implementation of mathematical theories. The Coq development of Leroy and Blazy is very close to the definitions presented in their paper. In fact, most of the definitions in their paper were transcribed directly from their Coq development, %. Statistically, their Coq development have
which has approximately 1070 lines of theorems and specifications, and 970 lines of proof scripts. 

Coq does not grant much support for proof automation. Therefore, Leroy and Blazy manually conducted most of the proofs. Yet, their proofs exhaustively use the omega tactic, a procedure that automates reasoning about equations and inequalities over the type \texttt{nat} of natural numbers. They also occasionally use other tactics (e.g.\ \texttt{eauto}, \texttt{congruence}), which %were claimed to be
the authors say were useful.

% \vspace{-0.5em}
% \paragraph{Challenge.}
% Most of their formalization was conducted in first-order logic. A question left open by the authors was whether their proofs could be automated using a verification framework for first-order-logic. 

% To answer the question Leroy and Blazy conducted a preliminary experience using Why (older version of Why3) with \texttt{Ergo}, \texttt{Simplify} \cite{Detlefs03simplify:a} and \texttt{Z3}. They only translated the axiomatics of the abstract memory model into Why (Section \ref{mem:axioms}), since at the time the recursive functions of the concrete memory model were hard to translate into Why's syntax. Their Why implementation generated 50 goals and the automatic theorem provers were able to prove 42 of them automatically. 

% We took the challenge and automated their whole formalization using the newest version of Why3.
\vspace{-0.5em}
\section{Our approach in Why3}
\label{sec:why3}
The formalization of Leroy and Blazy was conducted mostly in first-order logic: the authors use functions as data, but only to implement finite maps, which allows a first-order axiomatization. Henceforth, the definitions in Why3 are very similar to the fragment from Gallina (Coq's language). In fact, our development is very similar to the specification and theorems of the Coq version.

\paragraph{Implementation.} The Why version of Leroy and Blazy only contained the axiomatization of the abstract memory model. Our Why3 development have the axioms of the abstract memory model and additionally, the implementation of a concrete memory model, which allows an extraction of a correct-by-construction memory model. 
%
 %Why3 relies on external theorem provers, both automated and interactive, to discharge verification conditions. Our approach uses the following automatic theorem provers: \texttt{Alt-Ergo}, \texttt{Z3}, \texttt{CVC4} \cite{cvc4}, \texttt{SPASS} \cite{spass} and \texttt{Eprover}~\cite{eprover}.
%
Our formalization have approximately 1040 lines of theorems and specifications, roughly the same amount of the Coq version. 

\paragraph{Proofs.} The automatic theorem provers discharges automatically 114 from a total of 125 verification conditions using the auto level 3 tactic, which attempts to apply recursively four transformations (split\_all\_full, introduce\_premises, inline\_goal and split\_vc) and calls the provers with a larger time and memory limit (30 seconds and 4Gb of memory). Using the key 3 of the keyboard (shortcut for the auto level 3 tactic) we automatically prove the equivalent of approximately 870 lines of proof scripts in Coq (89.6\% of the total proof scripts lines). Appendix \ref{comparisonproof} shows the corresponding proofs in Coq and Why3 of one lemma of the 114.

For the remaining 9 verification conditions (VCs), we follow closely the Coq proof script, applying transformations one by one and calling the solvers for each transformation.
% and call the provers for each transformation of the Coq proof script
Three of the VCs require simple induction; four case analysis before applying the induction hypothesis; one additional properties in the context that we introduce with assertions and the last one a mix of induction, case analysis, assertions and additionally proves an existential statement (Table \ref{fig:prooftypes} summarizes the results). Yet, in all nine, the amount of transformations to complete the proof were significantly less than the steps of the proof in Coq (circa 117 transformations versus a total~of 220).

%quantificar
%tabelas
\vspace{-1em}
\subsubsection{General strategies to achieve a more automatic proof.}
\label{explicitinteractions}
The primary cons of Why3 regarding proofs with manually conducted interactions are: it is hard to have a concise global view of all transformations applied, especially on larger proof trees; moreover after making change in the code, most of the times Why3 is not able to propagate the exact transformations through the correct proof nodes, which quite often leads to losses of the whole proof session~\cite{preservinguserproofs}.

To circumvent this, we present the following general techniques to explicitly state the interactions required for the remaining 9 proofs as WhyML code. This allows the reader to have the essence of the whole proof just by looking at the source code. Bellow, we explain and illustrate the approaches namely in the treatment of induction, assertions and existential properties. The remainder 9 proofs use these techniques. These techniques can be used as general recipes for provers.

\paragraph{Induction.} Why3 provides a way to define lemma functions, which are special functions that serve not as actual code to execute but to prove the function's contract as a lemma. These are useful when proving properties by induction, as any recursive call to the functions means an application of the induction hypothesis. %has the ability to define recursive lemmas, which 
%allows the developer to simulate induction. 
Lets consider the following lemma:

\begin{why3}
lemma set_cont_outside:
    forall n, f:(int -> option (fset 'a, fset 'a)) , ofs i.
\end{why3}
We prove the lemma by induction on \texttt{n}. Let us now convert our lemma to a recursive lemma function. Recursive lemma functions are declared with the keywords \texttt{let rec lemma} and its parameters are the universally quantified variables in the normal lemma:

\begin{why3}
let rec lemma set_cont_outside (n: int) 
        (f: (int -> option (memtype, value))) (ofs i: int)
\end{why3}
Now we give a contract to the lemma, where each premise is now a precondition (declared with the \texttt{requires} keyword) and the conclusion a postcondition (declared with the \texttt{ensures} keyword). To guarantee the function terminates we state a variant, in this case \texttt{n} itself, which is the argument we are applying induction:
\begin{why3}
let rec lemma set_cont_outside (n: int) 
        (f: (int -> option (memtype, value))) (ofs i: int)
    requires { n >= 0 }
    requires { i < ofs \/ i >= ofs + n  }
    ensures { set_cont f ofs n i = f i }
    variant { n }
\end{why3}
Finally, we construct the correct recursive call by looking at the definitions of the functions within the postconditions. In this case we have the function \texttt{set\_cont}, which definition is as follows:

\begin{why3}
let rec ghost function set_cont (f: int -> content) 
        (ofs: int) (n: int) : int -> content
    requires { n >= 0 }
    variant { n }
= if n = 0 then f else set_cont (update ofs None f) (ofs + 1) (n-1)
\end{why3}
The recursive call of the lemma uses the same argument modifications of the recursive call of \texttt{set\_cont}:
\begin{why3}
let rec lemma set_cont_outside (n: int) 
        (f: (int -> option (memtype, value))) (ofs i: int) ...
= if n > 0 then set_cont_outside (n-1) (update ofs None f) (ofs + 1) i
\end{why3}
Why3 is now able to automatically generate a valid induction hypothesis and prove the lemma. The technique can be used as a general recipe for provers.

\paragraph{Asserts.} %Similar to Coq, in Why3 we can explicitly do assertions. For that, we need to convert the \texttt{lemma} to a \texttt{let lemma} (similar to what we did previously but without recursion) and 
%In the body of \texttt{let lemma}s, we introduce the assertion. 
We introduce assertions in the body of \texttt{let lemma}s. 
For example in the lemma \texttt{store\_mapped\_emb}, we assert that \texttt{b1} is a valid pointer in memory state \texttt{m1} and \texttt{b2} is a valid pointer in memory state \texttt{m2}:

\begin{why3}
let lemma store_mapped_emb 
    (val_emb: (int -> option (int, int)) -> value -> value -> bool) 
    (emb: int -> option (int, int)) (m1 m2: mem_) 
    (b1 ofs b2 delta: int) (v1 v2: value) (ty: memtype) (m1': mem_)
  requires { ... }
  ensures { ... }
= assert { valid_pointer_ ty m1 b1 ofs };
  assert { valid_pointer_ ty m2 b2 (ofs + delta) }
\end{why3}
The assertions are introduced in the hypotheses of the lemma and Why3 generates two new VCs to prove both the assertions are valid.

\paragraph{Witnesses for existential statements.} When one needs to prove an existentially quantified formula, sometimes the automatic theorem provers are not able to easily find the specific witness. Therefore, one needs to manually build it. Consider the following lemma: %\texttt{store\_lessdef}:

\begin{why3}
lemma store_lessdef:
  forall m1 m2 ty b ofs v1 m1' v2. mem_lessdef m1 m2 -> 
      store_ ty m1 b ofs v1 = Some m1' -> val_lessdef v1 v2 ->
      exists m2'. store_ ty m2 b ofs v2 = Some m2' /\ mem_lessdef m1' m2'
\end{why3}
We convert the \texttt{lemma} to a \texttt{let lemma} and as the lemma now needs to return a value, we change the signature of the function to return the type of the witness:

\begin{why3}
let lemma store_lessdef (m1 m2: mem_) (ty: memtype) (b ofs: int) 
    (v1: value) (m1': mem_) (v2: value) : (m2': mem_)
\end{why3}
We also need to change the postcondition, which should now ensure the same properties over the result of the function. This is achieved by replacing the existential quantified variable with the output value \texttt{m2'}:
\begin{why3}
ensures { store_ ty m2 b ofs v2 = Some m2' /\ mem_lessdef m1' m2'}
\end{why3}
Lastly, the function needs to return a proper witness that satisfied the postconditions. In this case, we manually build a new memory state with updated fields:
\begin{why3}
let lemma store_lessdef (m1 m2: mem_) (ty: memtype) (b ofs: int) 
    (v1: value) (m1': mem_) (v2: value) : (m2': mem_)
  ... = ...
  (mem_'mk (m2.nextblock) (m2.bounds_) (m2.freed)
    (update b (store_contents (m2.contents @ b) ty (ofs + 0) v2)
    (m2.contents)))
\end{why3}
The proof is now automatic and again, the technique is general.

\subsubsection{Outcome.} Applying these three techniques to the remainder 9 lemmas, we reduce 100\% the number of manually conducted transformations needed to complete the proofs. Table \ref{fig:beforeandafter} compares the before and after applying these techniques.

\begin{table}[ht]
    \centering
    \scalebox{0.85}{
    \begin{tabular}{|l|c|c|}
    \hline
     \multirow{2}{*}{\textbf{Lemma}} &  \multicolumn{2}{c|}{\textbf{Number of interactions (Why3)}} \\
    \cline{2-3}   & \hspace*{6mm} Before \hspace*{6mm}  & After \\ \hline
     1. check\_cont\_charact            &       7             &       0                  \\ \hline
     2. set\_cont\_outside              &       2             &         0           \\ \hline
     3. set\_cont\_inside               &       2               &       0      \\ \hline
     4. free\_list\_left\_emb           &       6       &           0       \\ \hline
     5. free\_list\_not\_valid\_block         &       6       &       0           \\ \hline
     6. embedding\_no\_overlap\_free\_list         &       6       &    0             \\ \hline
     7. free\_list\_fresh\_block         &       3       &              0    \\ \hline
     8. alloc\_list\_left\_inject         &       75       &          0       \\ \hline
     9. alloc\_list\_alloc\_inject         &       10       &         0       \\ \hline
    \end{tabular}
    }
    \medskip
    \caption{Number of transformations before and after}
    \label{fig:beforeandafter}

\end{table}

%Coq's language (\textit{Gallina}) and Why3's language (\textit{WhyML}) are both functional and quite similar (with little differences here and there). In short, the differences reside in the way that non-recursive and recursive function are declared.

% \begin{itemize}
%     \item In Coq non-recursive functions are defined with the keyword \texttt{Definition} and in Why3 with the keywords \texttt{let function};
%     \item Recursive function are defined with the keywords \texttt{Fixpoint} and \texttt{let rec function}, respectively;
%     \item Axioms are defined with the keywords \texttt{Hypothesis/Axiom} and \texttt{Axiom}, respectively.
% \end{itemize}
%Basically, one needs to adjust the declaration of the functions according to the corresponding language.
%Therefore, developing the memory model in Why3 is a straightforward process. One translates Coq implementation just following their definitions. 

\vspace{-2em}
\section{Proof Effort}
\label{sec:proofeffort}
\renewcommand{\arraystretch}{1.2}

Coq support for proof automation is limited, as one needs to explicitly apply a sequence of tactics and transformations to prove the desired goal. In contrast, one of the main focus of Why3 is automation, linking with SMT solvers to discharge proof obligations, what allows to develop proofs with less effort.

In this section we present metrics to compare our proofs with the Coq version. We used the original authors' proof as we are not aware of any improved version -- nowadays Coq has much more support to automation and experts may reduce significantly the amount of transformations.
Anyway, our goal is not a direct comparison between lines of proof scripts, but how much automation can be achieved, being this actually the original challenge left by Leroy and Blazy.

\vspace{-1em}
\subsubsection{Why3 fully automatic proof.}
All lemmas were automatically proved using the auto\_level\_3 tactic of Why3. The detailed proof results are available \href{https://htmlpreview.github.io/?https://gitlab.com/p.barroso/memory-model-c-why3/-/raw/explicit-induction/memorymodel-t/why3session.html}{here}. The table \ref{table:numberoflemmas} summarizes the structure of the proof.

%\vspace{-1.5em}
\begin{table}[ht]
\centering
\scalebox{0.52}{
\begin{tabular}{|l|ccccccc|c|}
\hline
\multirow{2}{*}{}             & \multicolumn{7}{c|}{\textbf{Module}}                                                                                                                                                                                                                                                                                            & \multirow{2}{*}{\textbf{Total}} \\ \cline{2-8}
                              & \multicolumn{1}{l|}{\textbf{Gen\_Mem\_Facts}} & \multicolumn{1}{l|}{\textbf{Ref\_Gen\_Mem\_Facts}} & \multicolumn{1}{l|}{\textbf{Concrete\_Mem}} & \multicolumn{1}{l|}{\textbf{Rel\_Mem}} & \multicolumn{1}{l|}{\textbf{Mem\_Extends}} & \multicolumn{1}{l|}{\textbf{Mem\_Lessdef}} & \multicolumn{1}{l|}{\textbf{Mem\_Inject}} &                                 \\ \hline
\textbf{Number of lemmas}     & \multicolumn{1}{c|}{7}                        & \multicolumn{1}{c|}{12}                            & \multicolumn{1}{c|}{61}                     & \multicolumn{1}{c|}{16}                & \multicolumn{1}{c|}{7}                     & \multicolumn{1}{c|}{6}                     & 15                                        & 124                            \\ \hline
\end{tabular}
}
\vspace{1em}
\caption{Number of lemmas per module}
\label{table:numberoflemmas}
\end{table}

%\vspace{-3em}
%\begin{table}[ht]
%\centering
%\scalebox{0.7}{
%    \begin{tabular}{|l|c|c|}
%        \hline
%        \multicolumn{1}{|c|}{Module} & Number of Lemmas & Automatically proved \\ \hline
%        Gen\_Mem\_Facts              & 7                & 7 (100\%)            \\
%        Ref\_Gen\_Mem\_Facts         & 12               & 12 (100\%)           \\
%        Concrete\_Mem                & 64               & 64 (100\%)           \\
%        Rel\_Mem                     & 17               & 17 (100\%)           \\
%        Mem\_Extends                 & 7                & 7 (100\%)            \\
%        Mem\_Lessdef                 & 7                & 7 (100\%)            \\
%        Mem\_Inject                  & 17               & 17 (100\%)           \\ \hline
%        Total                        & 131              & 131 (100\%)          \\ \hline
%    \end{tabular}
%}
%\end{table}
These results take into consideration the interactions stated explicitly in the code. Although the automatic theorem provers were able to prove all the lemmas automatically, we had to use the techniques in Section \ref{explicitinteractions} to aid the verification process. One might think these techniques do exactly the same thing as the transformations in the Coq proof scripts. In some sense that is true, we closely followed the Coq proof script and applied transformations one by one. However, when explicitly stating the transformations, one gives auxiliary results (e.g.\ auxiliary lemmas, assertions, concrete definitions) not a sequence of transformations/tactics as in Coq. On one hand, in Why3 we have the certainty that all auxiliary results are required to complete the proof (without them it is not possible to prove the lemmas). On the other, in Coq, one does not have the certainty that all the tactics are essential to the proof (one just knows that specific sequence completes the proof). Nonetheless, even without explicitly stating the interactive proofs, the effort to prove the lemmas in Why3 was way less than in Coq. Let us compare the non-fully automatic Why3 proof with the Coq version.

%the effort to prove the lemmas in Why3 was way less than in Coq. Let us compare the non fully automatic Why3 proof with the Coq version.
%quando tornamos explicita, não sabemos a sequencia
%mas os lemmas são muito mais informativos, tem um enunciado explicito (claro), enquanto que no Coq temos sequencia de taticas
%a prova precisa de pelo menos os resultados auxiliar definidos (se tirarmos algum a prova não é feita)
%no coq não sabemos se os passos anteriores são estritamente necessários

%\vspace{-0.5em}
\subsubsection{Why3 semi-automatic proof.}
The implementation and proofs without explicit interaction in the code can be consulted \href{https://gitlab.com/p.barroso/memory-model-c-why3/-/raw/poly/memorymodel.mlw}{here}. Even without stating explicitly the interactions, the theorem provers were able to prove 115 VCs automatically, which translates into a gain of automation circa 93\%. Table \ref{tab:numberoflemmas1} summarizes the results.

%\vspace{-1em}
\begin{table}[ht]
\centering
\scalebox{0.52}{
\begin{tabular}{|l|ccccccc|c|}
\hline
\multirow{2}{*}{}             & \multicolumn{7}{c|}{\textbf{Module}}                                                                                                                                                                                                                                                                                            & \multirow{2}{*}{\textbf{Total}} \\ \cline{2-8}
                              & \multicolumn{1}{l|}{\textbf{Gen\_Mem\_Facts}} & \multicolumn{1}{l|}{\textbf{Ref\_Gen\_Mem\_Facts}} & \multicolumn{1}{l|}{\textbf{Concrete\_Mem}} & \multicolumn{1}{l|}{\textbf{Rel\_Mem}} & \multicolumn{1}{l|}{\textbf{Mem\_Extends}} & \multicolumn{1}{l|}{\textbf{Mem\_Lessdef}} & \multicolumn{1}{l|}{\textbf{Mem\_Inject}} &                                 \\ \hline
\textbf{Number of lemmas}     & \multicolumn{1}{c|}{7}                        & \multicolumn{1}{c|}{12}                            & \multicolumn{1}{c|}{61}                     & \multicolumn{1}{c|}{16}                & \multicolumn{1}{c|}{7}                     & \multicolumn{1}{c|}{6}                     & 15                                        & 124                             \\ \hline
\textbf{Automatically proved} & \multicolumn{1}{c|}{7 (100\%)}                & \multicolumn{1}{c|}{12 (100\%)}                    & \multicolumn{1}{c|}{58 (95\%)}            & \multicolumn{1}{c|}{14 (87.5\%)}       & \multicolumn{1}{c|}{7 (100\%)}             & \multicolumn{1}{c|}{6 (100\%)}             & 11 (73.3\%)                               & 115 (92.7\%)                    \\ \hline
\end{tabular}
}
\medskip
\caption{Number of lemmas proved automatically}
\label{tab:numberoflemmas1}
\vspace{-2.5em}
\end{table}

\vspace{-0.5em}
%\begin{table}[ht]
%\centering
%\label{tab:numberoflemmas}
%\scalebox{0.7}{
%    \begin{tabular}{|l|c|c|}
%        \hline
%        \multicolumn{1}{|c|}{Module} & Number of Lemmas & Automatically proved \\ \hline
%        Gen\_Mem\_Facts              & 7                & 7 (100\%)            \\
%        Ref\_Gen\_Mem\_Facts         & 12               & 12 (100\%)           \\
%        Concrete\_Mem                & 64               & 61 (95.3\%)           \\
%        Rel\_Mem                     & 17               & 15 (88.2\%)           \\
%        Mem\_Extends                 & 7                & 7 (100\%)            \\
%        Mem\_Lessdef                 & 7                & 7 (100\%)            \\
%        Mem\_Inject                  & 17               & 13 (76.5\%)           \\ \hline
%        Total                        & 131              & 122 (93.1\%)          \\ \hline
%    \end{tabular}
%}
%\medskip
%\caption{Number of lemmas proved automatically}
%\end{table}

%\vspace{em}
These 115 VCs corresponds to a total of approximately 870 lines of the Coq proof script, which in turn consists of 89.6\% of the total lines of the entire Coq proof script. 
Table \ref{tab:numberoftransf} compares the amount of transformations used in interactive proofs in Why3 and Coq. One can see that in Why3 we have a reduction of 46.8\% of the transformations compared to the Coq version, which means significantly less proof effort.

%\vspace{-1em}
\section{Code extraction} 
One extra contribution of our work is the extraction of a correct-by-construction OCaml implementation of the memory model.

The implementation of Leroy and Blazy uses several undefined functions (e.g.\ \texttt{enough\_free\_memory}, \texttt{alignof}). In order to extract OCaml code, we need to organize the code involving these functions and insert them into a parameterized module (functor). For example, considering the function \texttt{enough\_free\_memory}, we include its signature in the scope of the following \texttt{Make} functor:

\begin{why3}
scope Make
    scope X
        val function enough_free_memory (mem: mem_) (i: int) : bool
    end = struct ... end
\end{why3}
With this manual transformation in the WhyML code (with the respect to the one we originally used for the proofs), the extraction becomes automatic in the sense that the obtained OCaml code compiles.

\vspace{-1em}
\section{Conclusions} %and Future Work
\label{sec:conclusions}
\vspace{-0.5em}
In this paper we present two versions in Why3 of the memory model of Leroy and Blazy, which is used in the Compcert compiler: from the proof point of view one is semi-automatic and the other results from an engineering effort to attain full automation (e.g.\ turning tactics into auxiliary results). Using the Why3 code extraction mechanism \cite{pereiraphd}, from both versions we get correct-by-construction OCaml implementations of the memory model.

The first version is a direct translation from the Coq development to Why3. We immediately gain 93\% of automation. The remaining 7\% were proved following closely the Coq proof script. Nonetheless, in Why3 the amount of transformations required for the theorem provers to validate the proofs were significantly less (117 transformations versus a total of 220). 

The second version is a fully automatic proof. We performed general techniques (explained in Section \ref{explicitinteractions}) to explicitly define the interactive proofs of the first version within the WhyML code. This improves readability, as one does not need to open the Why3 IDE to verify which transformations are used, it is possible to have the essence of the proof just by analyzing the code. Furthermore, as the proof is fully automatic, this approach also preserves the proof-sessions, allowing a simple replay to check the reproducibility of the results. 

\vspace{-1em}
\subsubsection{Lessons learned.}

Nothing of the memory model was inexpressible in WhyML. This confirms the premise that automatic verification tools are well-good candidates to defiant proofs of complex programs without losing expressiveness.

Furthermore, most of the proofs required a handful of transformations and nevertheless 93\% of the proof was performed by merely pressing a button (hence using automatic tactics, like \texttt{split}, etc). It is possible to save time and effort to the proofs that are actually challenging. 

It is undoubtedly that we are, at some point, limited to first order logic, however if one reaches the limits of these type of provers, one can still prove the remaining goals in Coq for example. In fact, Why3 provides a Coq tactic to call external theorem provers as oracles.

Nonetheless, to achieve an even higher level of automation we performed some refactoring to expose properties to the WhyML code: created auxiliary lemmas from patterns that constantly appear in the Coq proof script; defined inductive predicates to avoid proofs by induction; and transformed existential proofs to \texttt{lemma} functions which return the desired witness.

%Preserving, therefore, the proofs among code change

%Extraction
%Future work

\subsection*{Acknowledgements}
Work partially supported by the Portuguese Fundação para a Ciência e
Tecnologia via NOVA LINCS (UIDB/04516/2020) and by the first author PhD
grant (UI/BD/151265/2021).


\bibliographystyle{splncs04}
\bibliography{bib}

\appendix
%\newpage

\renewcommand{\thetable}{\Alph{section}\arabic{table}}
\setcounter{table}{0}

\section{Comparison of an automatic proof}
\label{comparisonproof}

\begin{minipage}{.48\textwidth}
Coq
\begin{lstlisting}[language=Coq]
Lemma store_mapped_emb:
    ...
Proof: (*@ \color{red}(166 transformations/tactics) @*)
  assert (valid_pointer ty m1 b1 ofs).
  eapply M.store_valid_pointer; eauto. 
  assert (valid_pointer ty m2 b2 (ofs + delta)).
  eapply valid_pointer_emb; eauto.
  destruct (M.valid_pointer_store _ _ _ _ v2 H6) 
    as [m2' STORE2].
  exists m2'; split. auto.
  red. intros ty' b1' ofs' v b2' delta' CP LOAD1.
  assert (valid_pointer ty' m1 b1' ofs').
  rewrite <- (store_valid_pointer_inv (...) ). 
  eapply M.load_valid_pointer; eauto.
  generalize (load_store_characterization (...) ).
  destruct (load_store_classification (...) );
  intro.
  ... (*@ \color{red} (+50 lines) @*)
Qed
\end{lstlisting}
\end{minipage} \hfill

\bigskip

\begin{minipage}{.48\textwidth}
Why3
\begin{why3tiny}
lemma store_mapped_emb:
  ...
\end{why3tiny}
\vspace{-1em}
\begin{figure}[H]
    \centering
    \hspace*{-4em}\includegraphics[width=4cm]{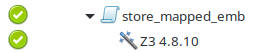}
\end{figure}

{
\small
\vspace{-1em}

\begin{itemize}
    \setlength{\itemindent}{-0.8em}
    \item The proof is immediate
    \item Proved in 0.15 seconds
\end{itemize}

}
\end{minipage}

\newpage

\section{Types of proofs}

\begin{table}[ht]
    \centering
    %\scalebox{1}{
    \begin{tabular}{|c|l|c|}
    \hline
    \multirow{2}{*}{\textbf{Type}} & \multirow{2}{*}{\textbf{Lemma}} & \textbf{Number of transformations} \\
    \cline{3-3} &  & Why3 \\ \hline
    {\color{orange} \scriptsize $\bullet$} & 1. check\_cont\_charact  &       7                   \\ \hline
    \color{orange} \scriptsize $\bullet$ & 2. set\_cont\_outside              &         2       \\ \hline
    \color{orange} \scriptsize $\bullet$ & 3. set\_cont\_insid            &       2      \\ \hline
    \color{blue} \scriptsize $\bullet$ & 4. free\_list\_left\_emb  &  6     \\ \hline
     \color{blue} \scriptsize $\bullet$ & 5. free\_list\_not\_valid\_block         &       6          \\ \hline
     \color{blue} \scriptsize $\bullet$ & 6. embedding\_no\_overlap\_free\_list  &    6             \\ \hline
     \color{blue} \scriptsize $\bullet$ & 7. free\_list\_fresh\_block       &              3    \\ \hline
     \color{red} \scriptsize $\bullet$ & 8. alloc\_list\_left\_inject &          75       \\ \hline
     \color{green} \scriptsize $\bullet$ & 9. alloc\_list\_alloc\_inject    &          10       \\ \hline
    \end{tabular}
    %}
    \medskip
    \begin{itemize}
    \footnotesize
    \item[\color{orange} \scriptsize $\bullet$] Simple induction
    \item[\color{blue} \scriptsize $\bullet$] Induction and case analysis
    \item[\color{green} \scriptsize $\bullet$] Assertions
    \item[\color{red} \scriptsize $\bullet$] The two above + \texttt{proof of an existential statement}
\end{itemize}
    \caption{Number of transformations per type of proof}
    \label{fig:prooftypes}
\end{table}

%\newpage
\section{Amount of transformations used in interactive proofs}
\vspace{-2em}

\begin{table}[H]
    \centering
    \begin{tabular}{|l|c|c|}
    \hline
    \multirow{2}{*}{\textbf{Lemma}} &  \multicolumn{2}{c|}{\textbf{Number of transformations}} \\
    \cline{2-3} & \hspace*{6mm} Coq \hspace*{6mm}  & Why3 \\ \hline
    1. check\_cont\_charact            &       26             &       7 (27\%)                  \\ \hline
    2. set\_cont\_outside              &       9             &         2 (22\%)           \\ \hline
    3. set\_cont\_inside               &       15               &       2 (13\%)      \\ \hline
    4. free\_list\_left\_emb           &       10       &           6 (60\%)       \\ \hline
    5. free\_list\_not\_valid\_block         &       16       &       6 (38\%)           \\ \hline
    6. embedding\_no\_overlap\_free\_list         &       11       &    6 (55\%)             \\ \hline
    7. free\_list\_fresh\_block         &       11       &              3 (27\%)    \\ \hline
    8. alloc\_list\_left\_inject         &       85       &          75 (88\%)       \\ \hline
    9. alloc\_list\_alloc\_inject         &       37       &          10 (27\%)       \\ \hline
    
    \textbf{Total}    & 220 & 117 (53.2\%) \\ \hline
    \end{tabular}
    \medskip
    \caption{Number of transformations in interactive proofs}
    \label{tab:numberoftransf}
\end{table}
\end{document}